\documentclass[12pt, centerh1]{article}

\usepackage{amsfonts}
\usepackage{natbib}
\usepackage{colonequals}
\usepackage{amsfonts} 
\usepackage{amsmath}
\usepackage{subfigure}
\usepackage{amssymb,lineno}
\usepackage[usenames,dvipsnames]{xcolor}
\usepackage{mathrsfs}
\usepackage{booktabs}
\usepackage{graphicx}
\usepackage{subfig}
\usepackage{multirow}
\usepackage{array}

\newcommand{\vecmu}{\mbox{\boldmath$\mu$}}

\title{Modelling Receiver Operating Characteristic Curves Using Gaussian Mixtures}
\author{Amay S.\ M.\ Cheam\thanks{Department of Mathematics and Statistics, University of Guelph, Guelph, Ontario, Canada, N1G~2W1. E-mail: acheam@uoguelph.ca} \ and Paul D.\ McNicholas}
\date{}

\begin{document}
\maketitle
%%%%%%%%%%
\begin{abstract}
%%%%%%%%%%
The receiver operating characteristic curve is widely applied in measuring the performance of diagnostic tests. Many direct and indirect approaches have been proposed for modelling the ROC curve, and because of its tractability, the Gaussian distribution has typically been used to model both populations. We propose using a Gaussian mixture model, leading to a more flexible approach that better accounts for atypical data. Monte Carlo simulation is used to circumvent the issue of absence of a closed-form. We show that our method performs favourably when compared to the crude binormal curve and to the semi-parametric frequentist binormal ROC using the famous LABROC procedure. 

\end{abstract}

%%%%%%%%%%%%
\section{Introduction}
\label{sec:Intro}
%%%%%%%%%%%%
The receiver operating characteristic (ROC) curve has gained tremendous popularity since its use in the signal detection theory during World War II. This phenomenon can be justified by the necessity to evaluate the performance of a diagnostic test, as noted by \cite{lusted71}. Despite being a useful tool to evaluate the efficiency of a diagnostic test, the ROC curve also presents a practical way to select an optimal threshold and to compare different tests. However, the empirical ROC curve is not desirable for the simple reason that it violates certain theoretical properties. Many authors have proposed different ways to model the ROC curve to circumvent this issue. Approaches to modelling the ROC curve within the literature can be divided into two categories: direct and indirect. 

The direct approach, which is less appealing, does not depend on any distributional hypotheses. The idea is to construct the ROC curve directly from the population scores, often in medical setting, are divided into two groups: diseased and non-diseased, without any assumptions \citep{lloyd98, zhou02}. As mentioned previously, the empirical ROC curve violates certain theoretical properties; e.g., it is not necessarily monotonic increasing. To overcome this obstacle, some authors proposed non-parametric estimation of the density function of each population using kernel smoothing methods \citep{hall03, lloyd98, lopez08, qiu01, zou97}. Hence, the problem is reduced to selection of an optimal bandwidth \citep{lloyd98, peng04, zhou02}. \cite{lloyd98} suggested using the bootstrap to minimize any distortion when smoothing the ROC curve. 

The indirect approach assumes that each population follows a certain distribution and implicitly derives a functional form for the ROC curve. To construct a curve, parametric and semi-parametric methods have been proposed. 
One of the parametric methods assumes that diseased and non-diseased populations follow a family of distributions such as: the Gaussian, which is the obvious and simple choice; the gamma \citep{dorfman97}; and others \citep{zweig93}. For the Gaussian assumption, \cite{goddard90} pointed out it is not always an adequate choice in some scenarios like prostate cancer. The authors emphasized that an inconsiderate and careless application of the method is not recommended, because it depends strongly on distributional assumptions. Furthermore, \cite{zhou02a} stressed the need to carefully verify the consistency of data with the assumptions. An alternative to the previous method is to specify a functional form of the ROC curve instead of assuming a distribution. For instance, both populations can be assumed to follow a logistic distribution with the same variance \citep{swets86}. \cite{england88} suggested an exponential model with two parameters. Both parametric methods are very similar because the distribution of the test scores entirely determines the shape of the ROC curve. The main advantages of a parametric method are simplicity, the smoothness of the curve, and an ability to work with a small number of parameters. 

The semi-parametric method is more attractive in terms of flexibility due to the presence of non-parametric and parametric components. The binormal model \citep{green66} is a good example; it assumes that both populations follow a Gaussian distribution after some monotone increasing transformation \citep{hanley96}. Hence, the problem is reduced to estimating the parameters, i.e., the slope and intercept. A range of solutions has been proposed using different techniques such as generalized least squares \citep{hsieh96}, maximum likelihood, pseudo-likelihood \citep{cai04,zhou08,zou00}, and others. For example, to obtain a smooth binormal ROC curve, \cite{metz98} developed an algorithm, called LABROC, which groups continuous data into a finite number of ordered categories and then uses the maximum likelihood algorithm from \cite{dorfman68} for ordinal data. A variation of this method was suggested by \cite{li99}, where they model the scores of a diagnostic test for non-diseased and diseased patients non-parametrically and parametrically, respectively. On the other hand, no functional relationship is assumed between these two distributions. Instead of directly modelling the distributions of the diagnostic scores of the two populations when the true status of the disease is known, another approach is to model the probability of knowing the disease status of the diagnostic scores using logistic regression \citep{qin03}. Evidently, like any estimation problem, the lack-of-fit can be an issue for the semi-parametric method. In addition to this estimation problem, the construction of confidence bands, for a given choice of both population distributions, is complicated.  

Our motivation is to develop a method that can give an estimate of the ROC curve with more flexibility and smoothness, produce reliable confidence bands, and ensure the natural monotonicity property of the ROC curve. We propose a Gaussian mixture (GM) distribution to model both non-diseased and diseased populations. This enables us to capture more complex behaviour and distribution shapes than the traditional normality assumption. By combining Monte Carlo simulation and the GM distribution, our method generates an ensemble of replica ROC curves and computes summary measures, such as the area under curve (AUC) and the partial AUC (pAUC), based on the ensemble. 

The remainder of the paper is organized as follows. In Section~\ref{sec:background}, we provide some background on ROC curves, followed by details of our proposed approach (Section~\ref{sec:method}). Results from simulation studies are provided in Section~\ref{sec:simulation} and real data analyses are discussed in Section~\ref{sec:real}. In Section~\ref{sec:conclusion}, some concluding remarks and possible extensions are discussed.  

%%%%%%%%%%%%%
\section{Background}
\label{sec:background}
%%%%%%%%%%%%%
The ROC curve is defined to be a plot of the true positive rate (TPR) against the false positive rate (FPR), or sensitivity versus $(1-\text{specificity})$, for various threshold values. This is generally a curve in the unit square anchored at $(0,0)$ and $(1,1)$, and above the line joining those points. Let $X \sim F$ and $Y \sim G$ be two independent continuous variables or two diagnostic variables coming from two populations, non-diseased and diseased, respectively. By convention, a patient is considered diseased if the value of the score is greater than a specified threshold. Note that we borrow the notation of \cite{gu08} in some of what follows. For a given threshold value $c_t \in \mathbb{R}$, 
	\begin{equation}
	\label{eq:FP}
		\text{FP}(c_t) = \int_{-\infty}^{+\infty}{ f_X(x)I\left( x-c_t \right)dx } = P(X > c_t),
	\end{equation}
	\begin{equation}
	\label{eq:TP}
 		\text{TP}(c_t) = \int_{-\infty}^{+\infty}{ g_Y(y)I\left( y-c_t \right)dy} = P(Y > c_t),
 	\end{equation}
where
 	\[ I(u)=
			\begin{cases}
				1,	&\text{if } u>0,\\
				0,	&\text{if } u\leq0.
			\end{cases}
	 \]	
Therefore, the ROC curve is obtained by
	 \begin{equation}
	 \label{eq:ROC}
	 	 \left\{(t,R(t))\right\} = \left\{(\text{FP}(c_t), \text{TP}(c_t))\right\},
	 \end{equation}	
where $t \in D \subset[0,1]$. 

When $t$ is given, $c_t = \bar{F}^{-1}(t)=F^{-1}(1-t)$, where $F^{-1}(\zeta)=\inf\{x:F(x)\geq\zeta\}$. If $\bar{F}^{-1}(t)$ exists, then the functional form of the ROC curve is given by
\begin{equation}
	\label{eq:ROCFF}
		R(t) = TP(c_t) = \bar{G}(\bar{F}^{-1}(t)) = \bar{G}(c_t)  = P(Y > c_t) = P(Y>\bar{F}^{-1}(t)),
	\end{equation}	
where $\bar{F}(u) = P(X>u)$ and $\bar{G}(u) = P(Y>u)$ are known as survival functions of $X$ and $Y$, respectively.  

The AUC is an extensively used summary index to quantify the information given by an ROC curve. The AUC, $A$, and its estimate $\hat{A}$ are defined as
\begin{equation}
	\label{eq:AUC}
		A = \int^{1}_{0}R(t) dt	\qquad 	\text{and}         \qquad	\hat{A} = \int^{1}_{0} \hat{R}(t) dt,
	\end{equation}
respectively, where $\hat{R}(t)$ is an estimate of $R(t)$.

%%%%%%%%%%%%%%
\section{Methodology}
\label{sec:method}
%%%%%%%%%%%%%%%
We refer to our approach, where a Gaussian mixture is used in conjunction with Monte Carlo simulation, as the MG method. The purpose of using the MG method is to produce a valid curve estimate and reliable confidence bands for any ROC curve. Using Gaussian mixtures leads to a more flexible model that accounts for data that might be considered atypical. Monte Carlo simulation is applied to circumvent the issue of the absence of a closed-form. Its properties enable the computation of confidence bands with ease. Because each pair $(X,Y)$ constructs one ROC curve, the idea of our MG method is to generate an ensemble of replica ROC curves by simulating many pairs $(X,Y)$, where $F$ and $G$ are assumed to be Gaussian mixture densities. Accordingly, we have
\begin{equation}
\label{eq:GM2}
	f(\bold{x}\mid\bold{\Theta}) = \sum^{K}_{k=1} \pi_{k} \phi(\bold{x}\mid \vecmu_k, \bold{\Sigma}_k),
\end{equation}
\noindent where
\begin{equation}
	\phi( \bold{x}\mid\vecmu_k, \bold{\Sigma}_k) = \frac{1}{ \sqrt{ (2\pi)^{p}|\bold{\Sigma}_k| } } \exp \left\{ -\frac{1}{2} (\bold{x} - \vecmu_{k})' \bold{\Sigma}^{-1}_{k} (\bold{x} - \vecmu_k) \right\}
\end{equation}
is the $k$th Gaussian component density with mean $\vecmu_k$ and covariance matrix $\bold{\Sigma}_k$, $\pi_k>0$ with $\sum_{k=1}^K\pi_k=1$ are the mixing proportions, and $\bold{\Theta} = (\pi_1, \ldots, \pi_K, \vecmu_1,\ldots,\vecmu_K,\bold{\Sigma}_1,\ldots,\bold{\Sigma}_K)$ is the collection of all model parameters. The density $g(\mathbf{y}\mid\bold{\Psi})$ is defined similarly, where $\bold{\Psi}$ is the collection of all model parameters.

The major difference between our parameter estimation approach and that of \cite{gu08} consists in the fact they use a Bayesian approach whereas we do not. They propose the Dirichlet process prior and then perform a bootstrap to resample. Like \cite{gu08}, we can lay out each step of our MG method, which combines Gaussian mixture modelling and Monte Carlo simulation. 

\textit{Step 1 (Parameter estimation for $F$ and $G$):}
Let $X$ and $Y$ be the vectors of scores of non-diseased and diseased populations, respectively.  Suppose both $X$ and $Y$ follow Gaussian mixture densities as in \eqref{eq:GM2}.  Parameter estimation is carried out via an expectation-maximization (EM) algorithm \cite{dempster77} and we thereby obtain $\bold{\Theta}_X$ and $\bold{\Theta}_Y$.

\textit{Step 2 (Generating the ensemble of random ROC curves):} 
After obtaining the parameter estimates, we generate  $(\tilde{X}, \tilde{Y})$ where $\tilde{X} \sim F(\bold{\Theta}_X)$ and $\tilde{Y} \sim G(\bold{\Theta}_Y)$. With the simulated ensemble, we compute  $\{(t,\tilde{R}(t))\}$ and $\tilde{A}$. This gives only one ROC curve, and after repeating this step $M$ times, via Monte Carlo simulation, we obtain 
\[\{(t,\tilde{R}_1(t))\},\ldots, \{(t,\tilde{R}_M(t))\}	\qquad	\text{and}	\qquad	\tilde{A}_1,\ldots,\tilde{A}_M.\]

\textit{Step 3 (Averaging the ensemble of random ROC curves):} 
The MG estimate, denoted as $\hat{R}^{MG}(t)$, is obtained by averaging the random realizations of the ROC curves such that $$\hat{R}^{MG}(t)=\text{mean}(\tilde{R}(t)),$$ where $t \in D \subset [0,1]$.  Similarly, we compute $$\hat{A}^{MG} = \int^{1}_{0} \hat{R}^{MG}(t) dt.$$

\noindent 
\textit{Remark 1} :
The estimate $\hat{R}^{MG}(t)$ is much smoother than the empirical estimate because it is obtained by averaging over the ensemble of random realizations $\tilde{R}(t)$.\\
\noindent 
\\
\textit{Remark 2} :
When plotting the curves, it is useful to add bands indicating the region where 95\% of the curves $\tilde{R}(t)$ lie. Therefore, to compute the confidence bands of the MG estimators of the ROC curves, we use the $M$ $\{(t,\tilde{R}(t))\}$ in Step~2. By the fundamental theory of Monte Carlo, which is based on the strong law of large numbers and the central limit theorem, we compute the MG standard error of $\hat{R}(t)$. For a given $t$, 
\begin{equation}
\label{eq:CI1}
	s_t = \sqrt{\frac{1}{M-1}\sum_{l=1}^{M} (\tilde{R}_{l}(t) - \hat{R}^{MG}(t))^2}.
\end{equation}
Hence, the upper and lower bounds of the $100(1-\alpha)$\% confidence interval can be written 
\begin{equation}
\label{eq:CI2}
	\left[\hat{R}^{MG}_{LB} (t),\ \hat{R}^{MG}_{UB} (t)\right]  = \left[\hat{R}^{MG}(t) - z_{1-\alpha} \frac{s_t}{\sqrt{M}},\ \hat{R}^{MG}(t) + z_{1-\alpha} \frac{s_t}{\sqrt{M}}\right]. 
\end{equation}

%%%%%%%%%%%%%%%%
\section{Simulation Studies}
\label{sec:simulation}
%%%%%%%%%%%%%%%%
In this section, we conduct several simulation studies to investigate the flexibility and the fit of the proposed MG approach by comparing it with existing procedures, i.e., the binormal model and the semi-parametric frequentist binormal ROC using the LABROC4 software \citep{webLABROC}. We assume that $X \sim\mathcal{N}({\vecmu}_X, {\bf \Sigma}_X)$ and $Y \sim \mathcal{N}({\vecmu}_Y, {\bf \Sigma}_Y)$. The parameters are chosen randomly. To evaluate the accuracy of MG method, we compute the AUC using the trapezoidal rule and the Mann-Whitney U test. 

\begin{figure}[!htp]
	\centering
		\includegraphics[width=14cm, height=6cm]{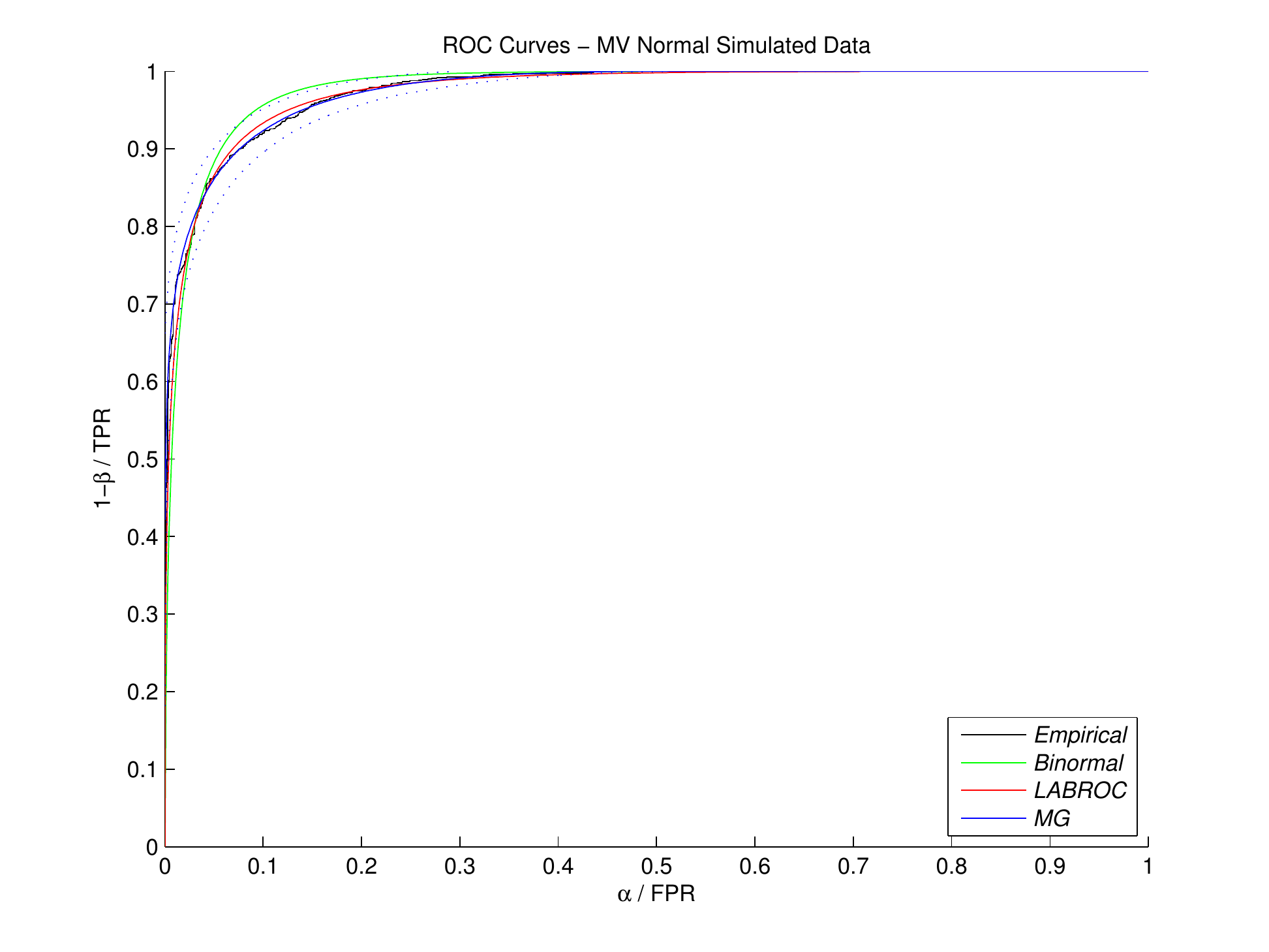}
		\caption{ROC curve for the simulated multivariate Gaussian data with strong discrimination (Figure~\ref{fig:Histo_MVNormalS}).}
		\label{fig:ROC_MVNormalS}
\end{figure}

\begin{figure}[!htp]
	\centering
		\includegraphics[width=12cm, height=6cm]{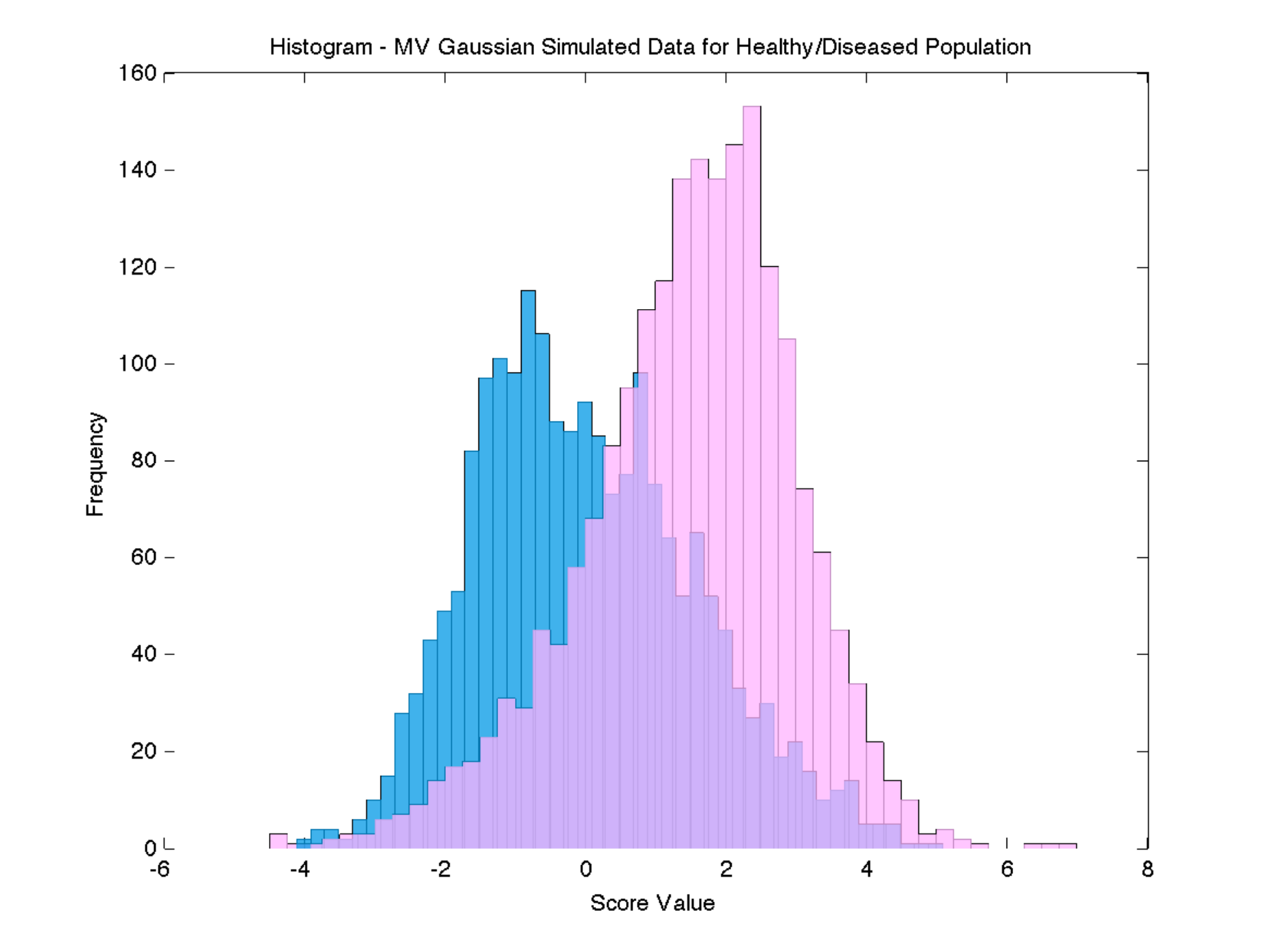}
		\caption{Histogram of the simulated multivariate Gaussian data with moderate discrimination.}
		\label{fig:Histo_MVNormalM}
\end{figure}

\begin{figure}[!htp]
	\centering
		\includegraphics[width=14cm, height=6cm]{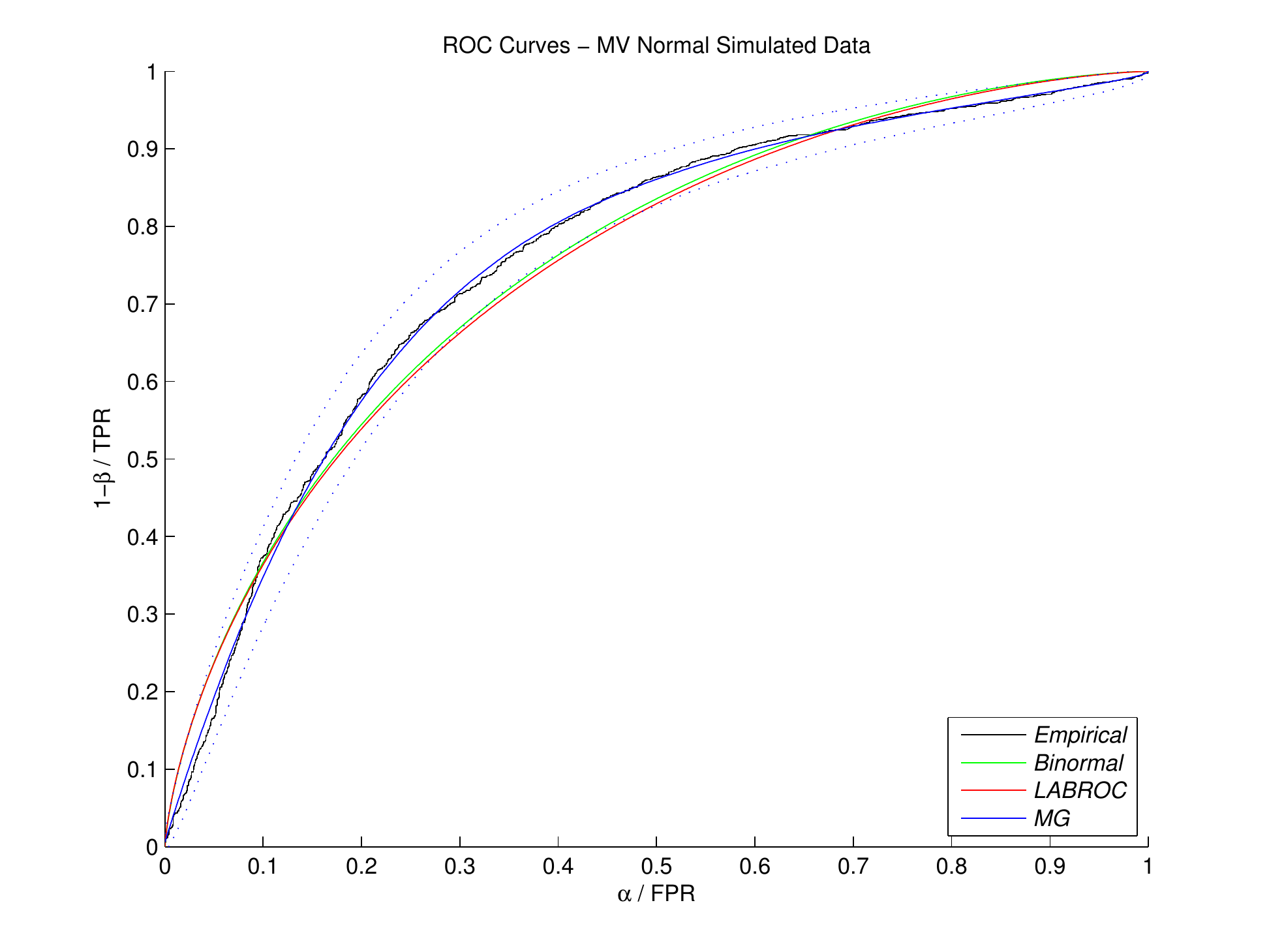}
		\caption{ROC curve for the simulated multivariate Gaussian data with moderate discrimination (Figure~\ref{fig:Histo_MVNormalM}).}
		\label{fig:ROC_MVNormalM}
\end{figure}

\begin{figure}[!htp]
	\centering
		\includegraphics[width=12cm, height=6cm]{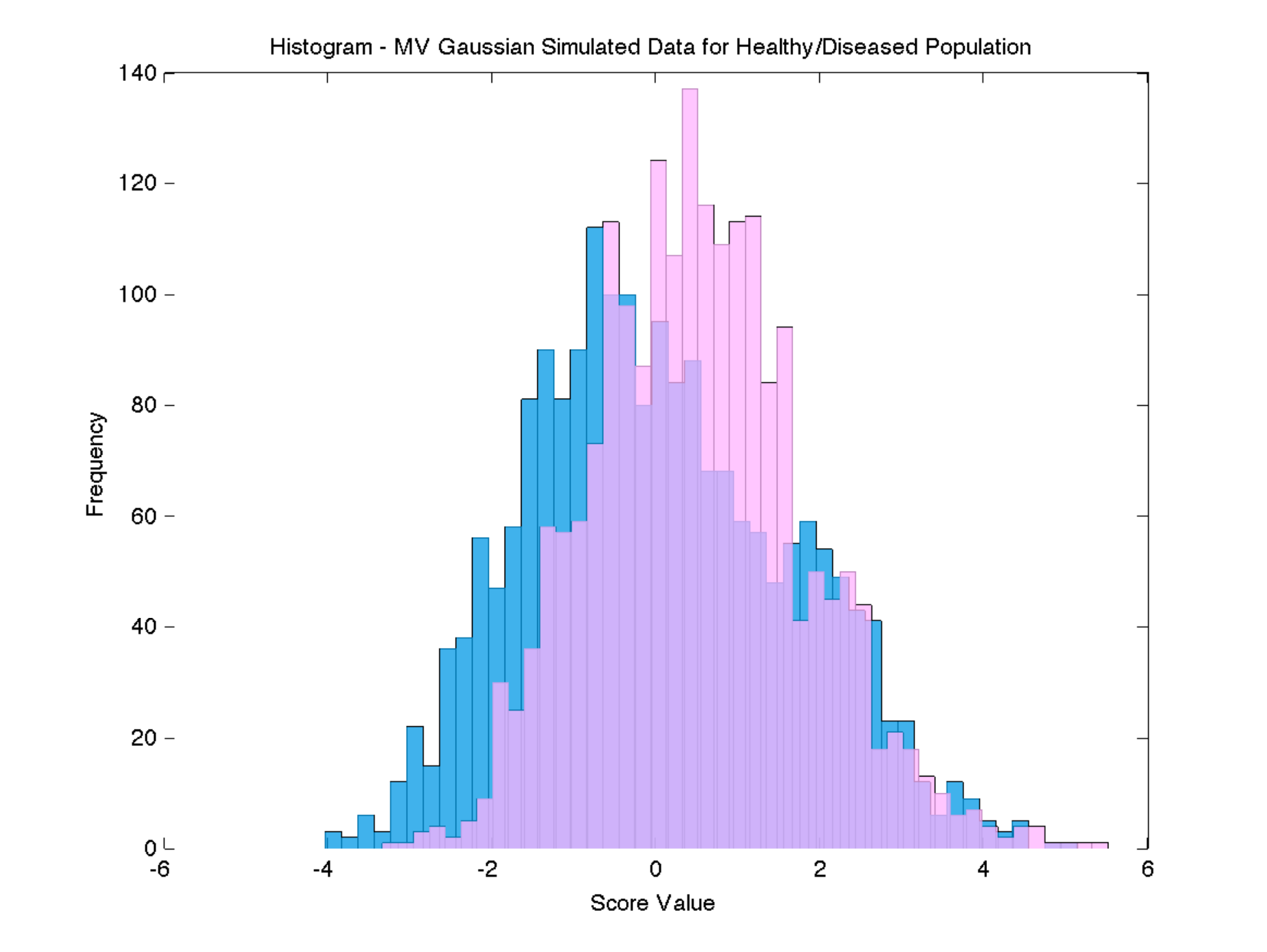}
		\caption{Histogram of the simulated multivariate Gaussian data with poor discrimination.}
		\label{fig:Histo_MVNormalP}
\end{figure}

\begin{figure}[!htp]
	\centering
		\includegraphics[width=14cm, height=6cm]{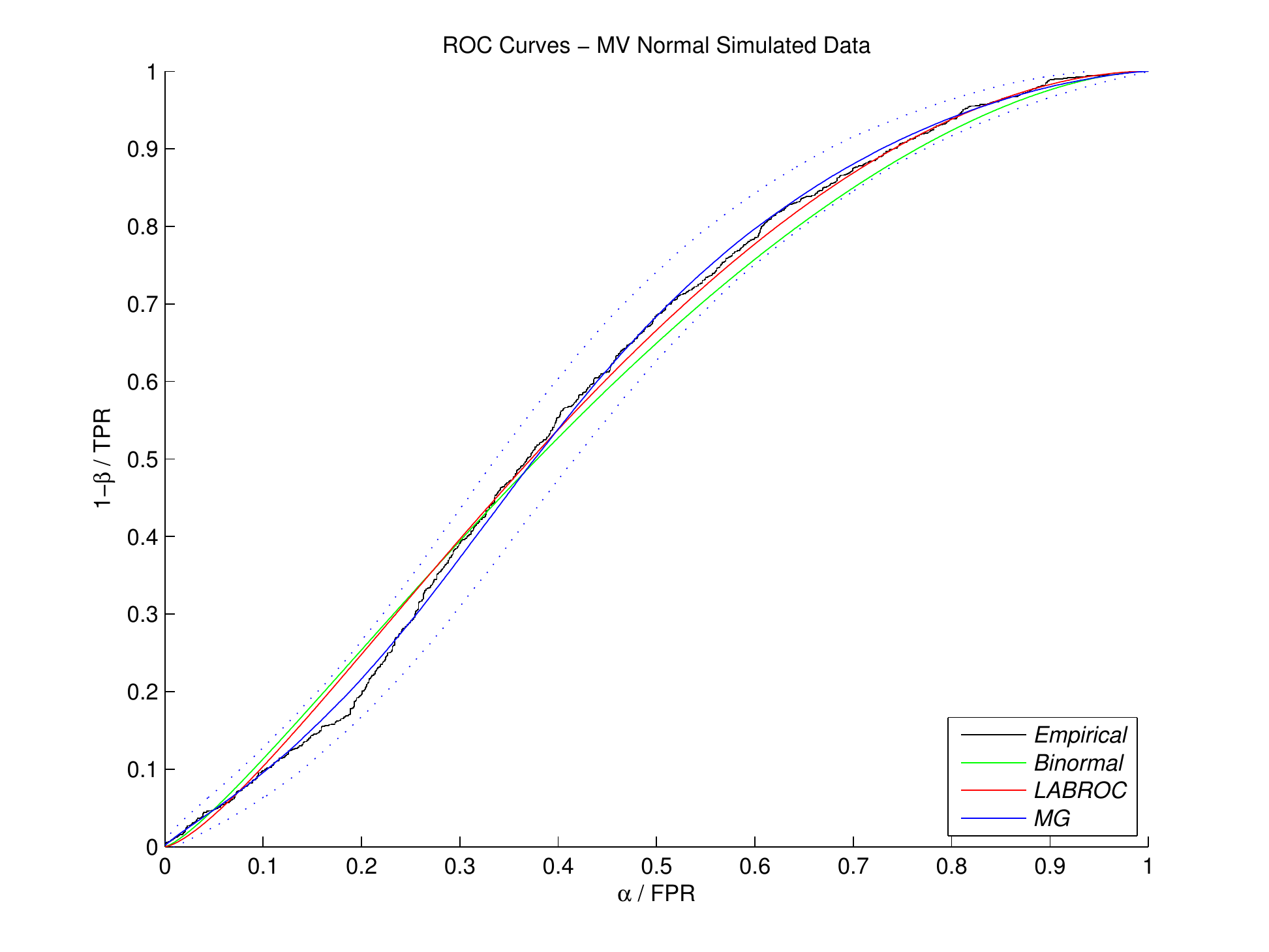}
		\caption{ROC curve for the simulated multivariate Gaussian data with poor discrimination (Figure~\ref{fig:Histo_MVNormalP}).}
		\label{fig:ROC_MVNormalP}
\end{figure}

To visualize the flexibility of the MG method, three cases of discrimination are examined: strong, moderate, and poor. From Figure~\ref{fig:Histo_MVNormalS}, we observe that the diseased population is a bimodal distribution. This scenario is practically relevant because the diseased population may contain a subpopulation of patients at different stages of a disease. Our goal is to replicate the empirical curve but with more smoothness. When strong discrimination is present (Figure~\ref{fig:ROC_MVNormalS}), we can observe that the MG curve performs as well as the commonly used LABROC method and significantly better than the crude binormal curve. Furthermore, for the two other cases, we notice that the MG curve follows the empirical curve closely when compared to the binormal and the LABROC approaches (Figures~\ref{fig:Histo_MVNormalM}--\ref{fig:ROC_MVNormalP}). The bands (dashed lines) in these graphs indicate the region covering 95\% of our simulated curves; recall that the MG curve represents an averaging of these simulated curves. The crude binormal curve is obtained directly from the following equation without any monotonic transformation:
\begin{equation}
\label{eq:ROCN}
	R(t) = \Phi(a + b\Phi^{-1}(t)),
\end{equation}
where
\begin{equation*}
\label{eq:ab}
	a = \frac{\mu_D-\mu_N}{\sigma_D}	\qquad	\text{and}		\qquad	b = \frac{\sigma_N}{\sigma_D}.
\end{equation*}
It follows that the AUC of a binormal curve is given by
\begin{equation}
\label{eq:AUC}
	A = \Phi \left( \frac{a}{\sqrt{1+b^2}} \right).
\end{equation}
\begin{figure}[!htp]
	\centering
		\includegraphics[width=12cm, height=6cm]{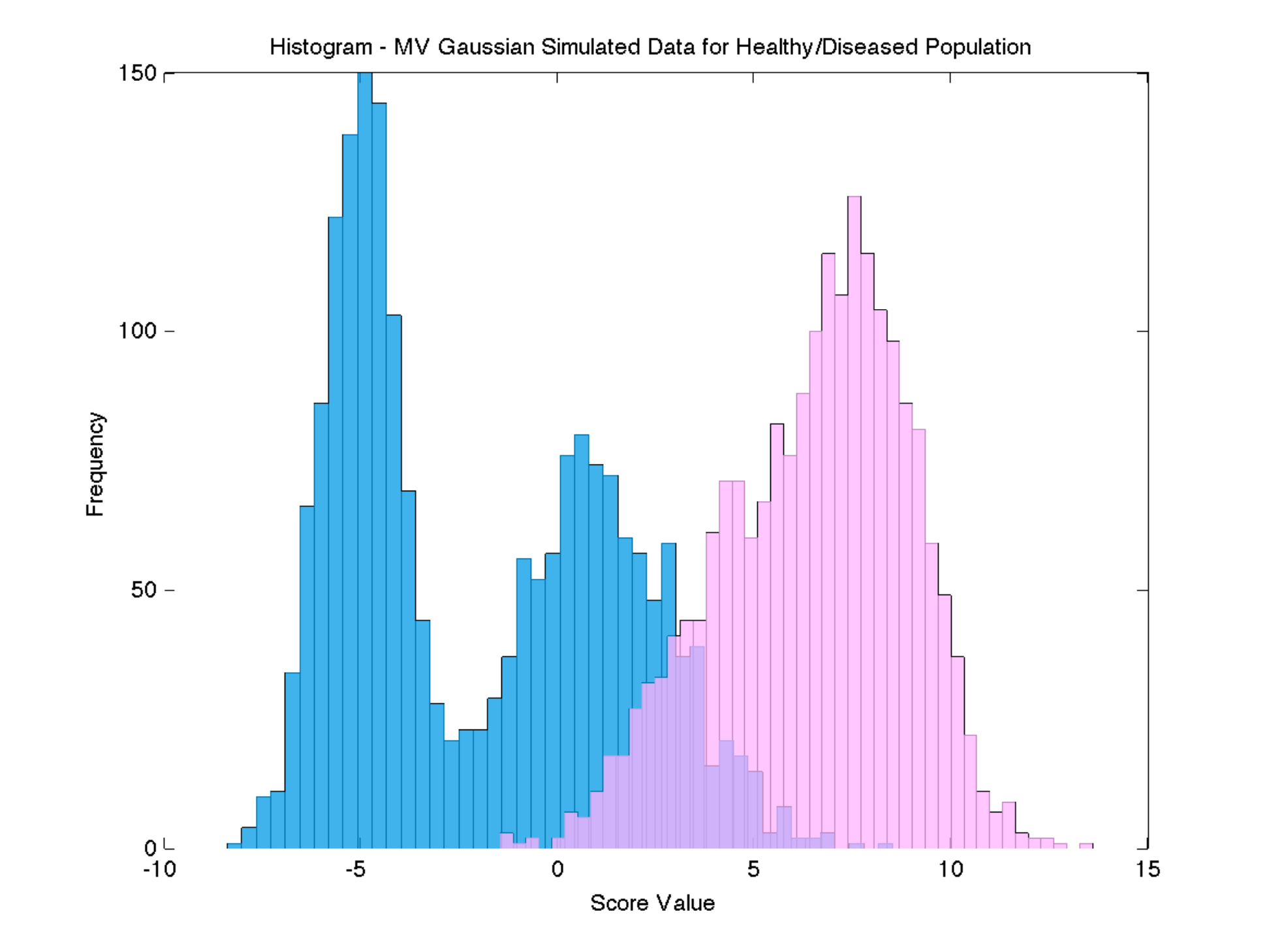}
		\caption{Histogram of the simulated multivariate Gaussian data with strong discrimination.}
		\label{fig:Histo_MVNormalS.pdf}
\end{figure}


Another way to compare the MG method against the two binormal methods is to calculate a summary measure such as the AUC (Table~\ref{table:SimTab}). We observe that our MG method obtained an AUC close to the empirical AUC and performed relatively well compared to the binormal and LABROC procedures. These simulation studies show that our approach is at least as accurate as the two classical binormal methods and sometimes superior.   

\begin{table}[!htp]
	\caption{AUC values for the simulated multivariate Gaussian data, with the model closest to the empirical curve in bold font.}
	\centering
		\begin{tabular*}{\textwidth}{@{\extracolsep{\fill}}llcc}
		\hline
		  {Level of}	&	{ROC Model} & \multicolumn{2}{c}{{AUC}} \\ \cline{3-4}
			{Discrimination} &	 & {Trapezoidal} & {Mann-Whitney} \\
		\hline
								 &	Empirical &	0.9760 &	0.9759 \\
			Strong	   			 &	Binormal & 	0.9787 &	0.9787* \\
								 &	LABROC &       0.9750  &  0.8337 \\
								 &	MG 	        &	{\bf 0.9764} &	{\bf 0.9757} \\
		\hline
								 &	Empirical &	0.7599 &	0.7597 \\
			Moderate	 			 &      Binormal & 	0.7521 &	0.7521* \\
								 &	LABROC &	0.7479 &  0.7163 \\
								 & 	MG	        & 	{\bf 0.7593} &	{\bf 0.7588} \\
		\hline
								 &  	Empirical &  	0.6008 &	0.6006 \\
			Poor					 &  	Binormal &   	0.5550 &	0.5950* \\
								 &	LABROC &	0.6024 &  0.6053 \\
								 &  	MG	        &   	{\bf 0.6013} &	{\bf 0.6011} \\
		\hline
		\multicolumn{ 2}{l}{\scriptsize{*Using the binormal crude AUC equation}} &            &            \\
		\end{tabular*}
	\label{table:SimTab}
\end{table}

%%%%%%%%%%%%%%%%%
\section{Real Data Analyses}
\label{sec:real}
\begin{figure}[!htp]
	\centering
		\includegraphics[width=12cm, height=6.5cm]{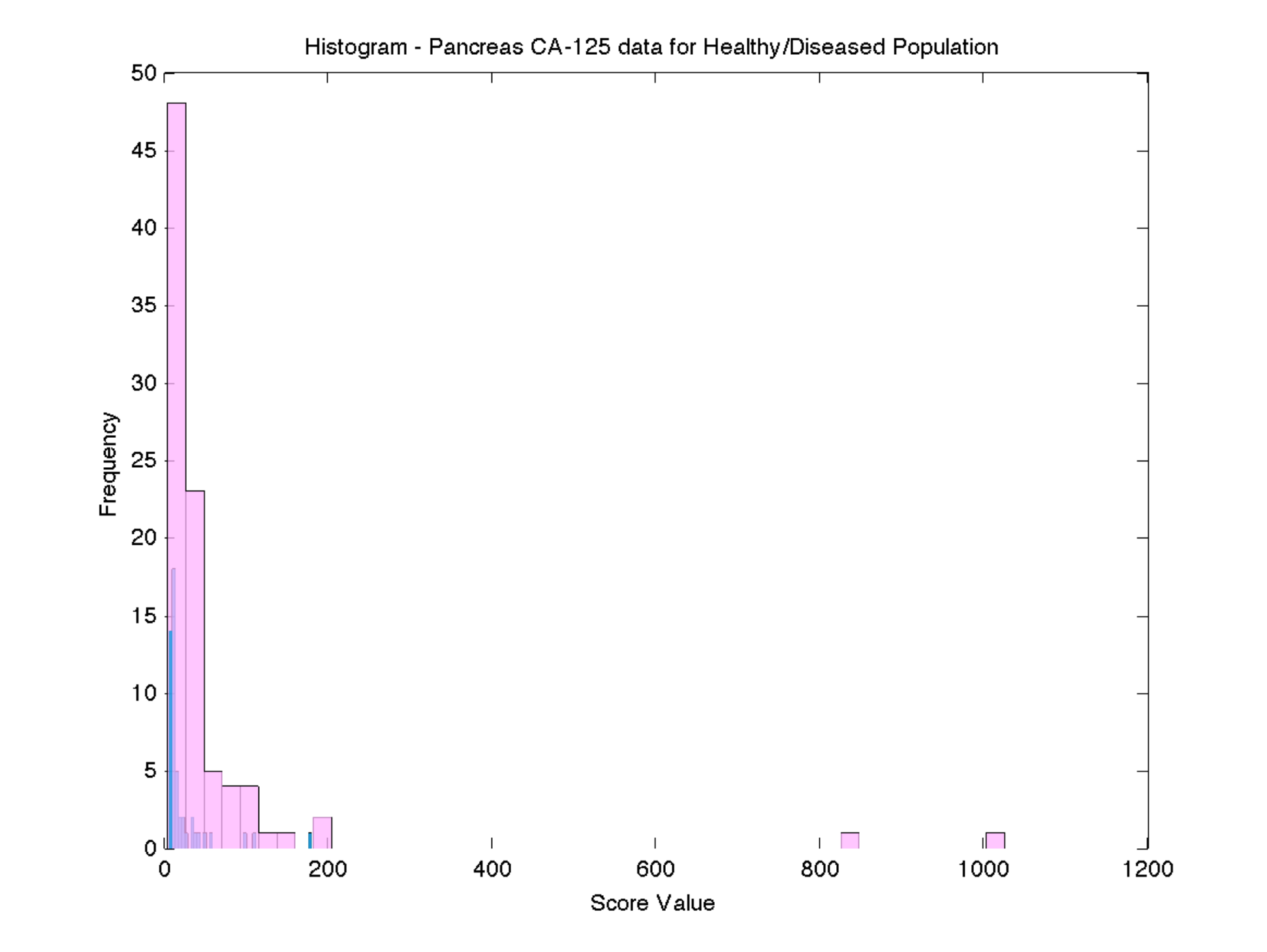}
		\caption{Histogram of the pancreatic cancer data using biomarker CA 125.}
		\label{fig:Histo_PancreasCA125}
\end{figure}

\begin{figure}[!htp]
	\centering
		\includegraphics[width=14cm, height=6.5cm]{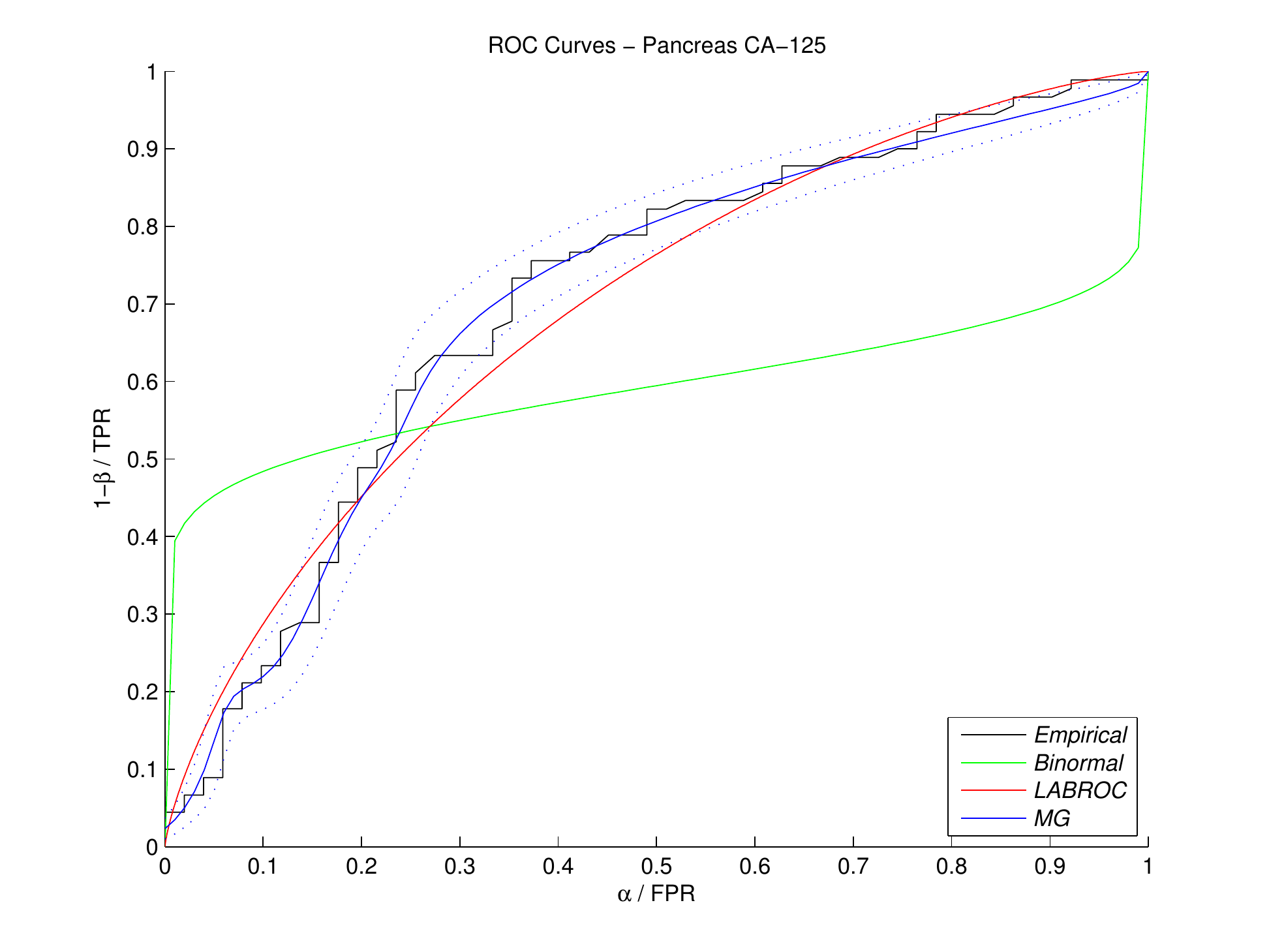}
		\caption{ROC curve for the pancreatic cancer data using biomarker CA 125.}
		\label{fig:ROC_PancreasCA125}
\end{figure}

\begin{figure}[!htp]
	\centering
		\includegraphics[width=12cm, height=6.5cm]{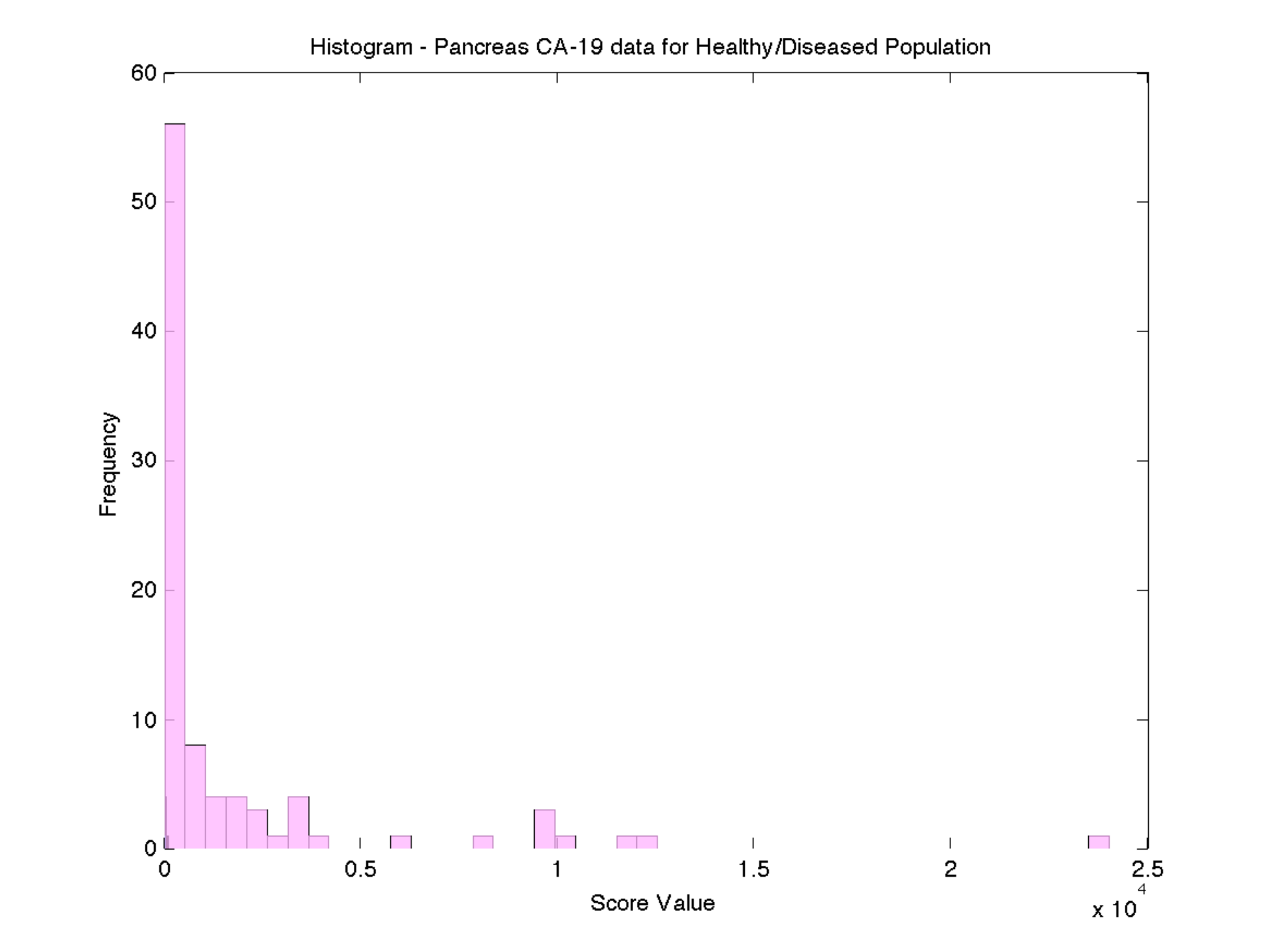}
		\caption{Histogram of the pancreatic cancer data using biomarker CA 19-9.}
		\label{fig:Histo_PancreasCA19}
\end{figure}

\begin{figure}[!htp]
	\centering
		\includegraphics[width=14cm, height=6.5cm]{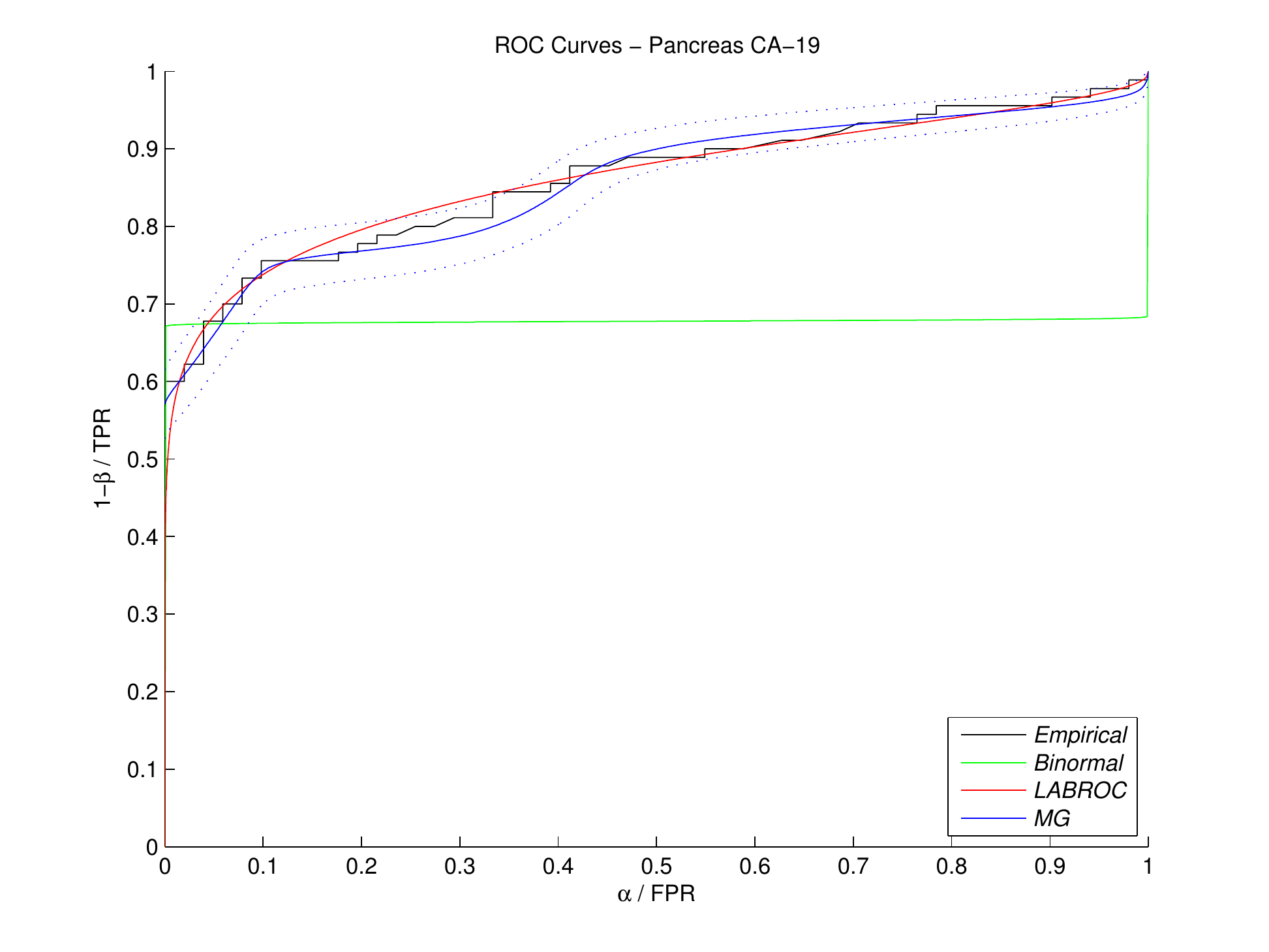}
		\caption{ROC curve for the pancreatic cancer data using biomarker CA 19-9.}
		\label{fig:ROC_PancreasCA19}
\end{figure}

%%%%%%%%%%%%%%%%%
Having observed favourable results on the simulation studies, we illustrate our newly proposed method on publicly available case-control cancer data published by \cite{wieand89}. This data set has been used extensively in the literature to illustrate newly developed methods for ROC curves; accordingly, we selected it to compare our method to the current state-of-the-art. This study examined two biomarkers: a cancer antigen (CA 125) and a carbohydrate antigen (CA 19-9). The data consist of 90 selected cases representing patients with pancreatic cancer as well as 51 controls without cancer but with pancreatitis. We used the two biomarkers to illustrate the application of our methodology. From Figures~\ref{fig:ROC_PancreasCA125} and~\ref{fig:ROC_PancreasCA19}, we observe that our MG method undoubtedly outperforms the crude binormal approach. The poor performance of the binormal curve can be explained by the unsuitability of the normality assumption for these data (see Figures~\ref{fig:Histo_PancreasCA125} and~\ref{fig:Histo_PancreasCA19}). Compared to the LABROC procedure, our MG method performs relatively well in terms of replication and closeness to the empirical curve. Without any monotonic transformation, the MG method outperforms the crude binormal ROC and performs as well as LABROC. As suspected, the AUC of our approach is closer to the empirical than the binormal for both biomarkers (Table~\ref{table:RealTab}).  For CA 125, our MG method obtains a summary index closer to the empirical curve than the LABROC. For biomarker CA 19-9, LABROC gives a better AUC using the trapezoidal rule; however, when using the Wilcoxon-Mann-Whitney statistic, our MG method performs better.
\begin{table}[!h]
	\caption{AUC values for pancreatic cancer data using two biomarkers, with the model closest to the empirical curve in bold font.}
	\centering
		\begin{tabular*}{\textwidth}{@{\extracolsep{\fill}}llcc}
		\hline
		  {Biomarker}	&	{ROC Model} & \multicolumn{2}{c}{{AUC}} \\ \cline{3-4}
			{} &	{} & {Trapezoidal} & {Mann-Whitney} \\
		\hline
						&  Empirical &     0.7143 	&     0.7056 \\
			Pancreas		&  Binormal  &	  0.5924	&     0.5924* \\
			 CA 125		&  LABROC  &	  0.6946    &     0.6808 \\
						&  MG	    &     {\bf 0.7147} &     {\bf 0.7143} \\
		\hline
						&  Empirical &     0.8651	&     0.8614 \\
			Pancreas		&  Binormal  &	  0.6774	&     0.6776* \\
			 CA 19-9		&  LABROC  &   {\bf 0.8625}    &     0.7793 \\
						&  MG	    &    0.8569     &    {\bf 0.8565} \\
		\hline
		\multicolumn{ 2}{l}{\scriptsize{*Using the binormal crude AUC equation}} &            &            \\
		\end{tabular*}
		\label{table:RealTab}
\end{table}


%%%%%%%%%%%%%%%%%
\section{Concluding Remarks}
\label{sec:conclusion}
%%%%%%%%%%%%%%%%%
In this paper, we have outlined a methodology to estimate the ROC curve with more flexibility and smoothness than provided by existing approaches. The proposed method utilizes the Gaussian mixture in conjunction with Monte Carlo simulation, and we refer to our approach as the MG method. The performance of the MG method was illustrated via several simulation studies and real data on pancreatic cancer. We found that our MG curve performed favourably when compared to the crude binormal curve in term of flexibility and fitting. Even without any monotonic transformation, the MG produced similar, if not better, results than the LABROC procedure. Furthermore, the MG method does not require any assumptions other than that the populations follow a Gaussian mixture. An interesting avenue for future work is to extend this approach using a non-Gaussian mixture model instead of a Gaussian mixture model. 

%%%%%%%%%%%%
%	REFERENCES	    %
%%%%%%%%%%%%
%\bibliographystyle{chicago}
%\bibliography{Biblio_S2013}

\end{document}